# A Photonic Integrated Circuit based Compressive Sensing Radio Frequency Receiver Using Waveguide Speckle

**David B. Borlaug,[1*] Steven Estrella,[2] Carl Boone,[1]**
**George A. Sefler,[1] Thomas Justin Shaw,[1] Angsuman Roy,[2]**
**Leif Johansson,[2] George C. Valley[1]**

[1]*The Aerospace Corporation, 2310 E. El Segundo Blvd., El Segundo, CA 90245, USA*
[2]*Freedom Photonics, 41 Aero Camino, Santa Barbara, CA 93117, USA*
**david.b.borlaug@aero.org*

**Abstract:** A photonic integrated circuit (PIC) comprised of an 11 cm multimode speckle waveguide, a 1x32 splitter, and a linear grating coupler array is fabricated and utilized to receive 2 GHz of radio-frequency (RF) signal bandwidth from 2.5 to 4.5 GHz using compressive sensing (CS). Incoming RF signals are modulated onto chirped optical pulses which are input to the multimode waveguide. The multimode waveguide produces the random projections needed for CS via optical speckle. The time-varying phase and amplitude of two test RF signals between 2.5 and 4.5 GHz are successfully recovered using the standard penalized $l_1$-norm method. The use of a passive PIC serves as an initial step towards the miniaturization of a compressive sensing RF receiver.

## 1. Introduction

Compressive sensing (CS) enables the capture of multiple narrow-band signals that sparsely occupy broad bandwidth domains while sampling below the Nyquist rate. In this work, 16 radio-frequency (RF) channels, each with an effective sample rate of 35 MSps, are used to recover RF signals across 2 GHz of RF bandwidth from 2.5 to 4.5 GHz. Compared with direct Nyquist sampling which requires 4 GSps, the CS system required 4000 / (35 x 16) = 7 times fewer recorded samples and a maximum sampling rate 4000 / 35 = 114 times smaller.

Previous works have demonstrated RF signal recovery [1], [2], [3], [4] and optical reservoir computing [5] using the speckle of a multimode optical fiber. The speckle of a multimode waveguide in a photonic integrated circuit (PIC) was previously utilized in demonstrating an RF spectrometer [4], [6]. Other workers in the field have demonstrated compressive sensing in the optical domain without the use of speckle. In those schemes a pseudo-random bit sequence is generated electronically to act as the measurement matrix. The sequence is modulated onto a chirped optical pulse containing RF modulation, thereby performing the compressive sensing matrix multiplication during modulation, without the use of passive speckle [7], [8]. In this paper, we demonstrate for the first time the use of waveguide speckle for the compressive sensing of RF signals in a compact PIC. The demonstration serves to identify a pathway toward a fully integrated optoelectronic speckle-based RF compressive sensing system.

The paper is organized as follows: compressive sensing is motivated for RF signal reception using optical speckle; the theoretical ability of multimode waveguides to produce speckle is investigated, and the baseline speckle waveguide design is introduced; the development of a PIC for radio frequency reception using waveguide speckle is detailed, including its various integrated components; the experimental compressive sensing setup is discussed, results are presented, and the paper is concluded.

## 2. Compressive Sensing of RF Signals -- A Speckle-Based Measurement Matrix

In traditional sampling, the required sampling rate is determined by the signal's Nyquist rate. In compressive sensing, the required sampling rate is determined by the information content and sparsity of the signal. An example of a sparse signal is multiple narrow-band signals that sparsely occupy a broad bandwidth domain. Sampling well below the broad bandwidth's Nyquist rate reduces the need for data storage, reduces data-payload transmission time, and reduces overall receiver size, weight, and power [3], [9].

Compressive sensing maps sparse information from a high-dimensionality space into a lower-dimensionality space while retaining the salient information. In compressive sensing, an input signal vector $\vec{x}$ (of large dimensionality $n \times 1$) is recovered from a measurement signal vector $\vec{y}$ (of small dimensionality $m \times 1$) where $m << n$. Compressive sensing measurements record $\vec{y}$, which is the result of the input signal $\vec{x}$ being multiplied by a measurement matrix $\Phi$.

Eqn. 1 $$\vec{y} = \Phi\vec{x}$$

The goal of compressive sensing is to efficiently recover $\vec{x}$ from $\vec{y}$. Here, the measurement matrix $\Phi$ has dimensionality $m \times n$. Eqn. 1 can be recast to investigate the presence of a sparse basis set upon which $\vec{x}$ can be mapped.

Eqn. 2 $$\vec{y} = \Phi\vec{x} = \Phi\Psi^{-1}\vec{s} = \Theta\vec{s}$$

Here $\vec{s} = \Psi\vec{x}$ is the sparse representation of $\vec{x}$ and $\Theta = \Phi\Psi^{-1}$. $\vec{x}$ can be recovered from $\vec{y}$, if $\Psi$ can be determined such that $\vec{s}$ is sparse. $\vec{s}$ is sparse if $\vec{s}$ has $K$ non-zero elements, $K \ll N$, and $m$ is somewhat greater than $K$. In other words, $\Psi$ is a transform that maps $\vec{x}$ into a sparse representation $\vec{s}$, [3] [9], [10], [11], [12], [13], [14]. Additionally, the measurement matrix $\Phi$ must be incoherent with basis $\Psi$ in which signal $\vec{x}$ has sparse representation, [9].

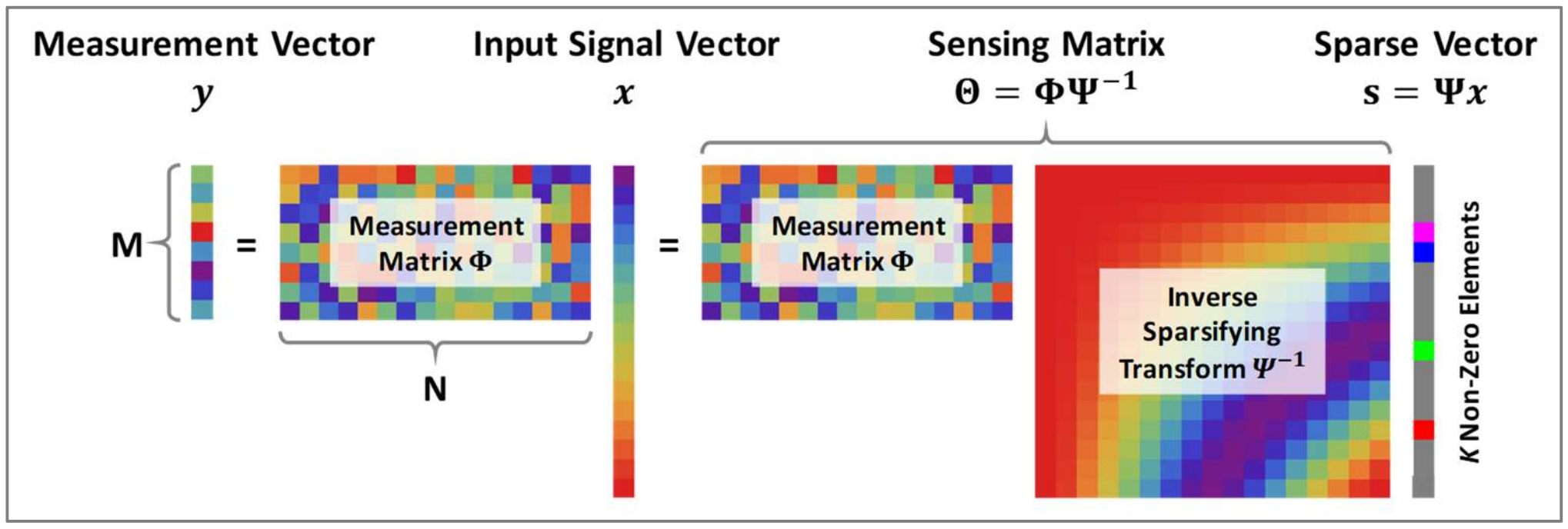

Fig. 1: Compressive sensing enables the recovery of a sparse input signal vector from a reduced dimensionality measurement vector.

A key challenge for CS systems is finding a practical way to efficiently perform the $\Phi\vec{x}$ or $\Theta\vec{s}$ matrix multiplication in real time. This work performs the matrix multiplication in the analog optical domain using waveguide speckle. Waveguide speckle is an inherent property of passive multimode waveguides. Matrix multiplication using speckle requires no active components or energy expenditure. The process begins by transcoding the input time series waveform on a chirped optical pulse. Next the pulse is input to the multimode waveguide where the waveguide's pseudo-random speckle pattern serves as the measurement matrix and performs the matrix multiplication. Because CS recovery calculations require accurate knowledge of $\Phi$ or $\Theta$ [15], $\Phi$ or $\Theta$ must be reproducible and amenable to calibration. After detecting the pseudo-random speckle pattern $\vec{y}$, the input time time series $\vec{x}$ can be efficiently recovered using the penalized $l_1$-norm [9]. Other methods such as orthogonal matching pursuit may similarly be utilized [16], [17].

## 3. Speckle in Planar Waveguides

In this section the speckle of a multimode planar dielectric waveguide is theoretically and experimentally investigated. Additionally, the silicon-based speckle mixer design used in the rest of this manuscript is introduced. The theoretical electric field profile of a rectangular dielectric slab waveguide is written below [2], [18].

Eqn. 3 $$E_Y(x,z,n,f) = \mathrm{A}\left\{cos[h(n,f)\cdot x] - \frac{q(n,f)}{h(n,f)} sin[h(n,f)\cdot x]\right\} e^{i\beta(n,f)z}$$

Eqn. 4 $$h(n,f) = \sqrt{n_1(f)^2 k^2 - \beta(n,f)^2}$$

Eqn. 5 $$q(n,f) = \sqrt{\beta(n,f)^2 - n_2(f)^2 k^2}$$

Eqn. 6 $$\tan[h(n,f)d] = \frac{2q(n,f)}{h(n,f)\left(1-\frac{q(n,f)^2}{h(n,f)^2}\right)}$$

Here $A$ is the field amplitude, $n$ is the integer mode number, $n_1(f)$ is the waveguide refractive index, $n_2(f)$ is the cladding refractive index, $k = 2\pi/\lambda = 2\pi f/c$ is the free-space wavenumber, d is the waveguide width, $\beta(n,f)$ is waveguide propagation constant, $h(n,f)$ is the wavenumber of the transverse mode profile in the guided core region, and $q(n,f)$ is the wavenumber of the transverse mode profile in the unguided cladding region. Light propagates in the z-direction and the refractive index contrast is witnessed in the x-direction only.

To produce solutions for Eqn. 3, $h(n,f)$, $q(n,f)$, and $\beta(n,f)$ must be simultaneously solved for using the transcendental forms of Eqn. 4, Eqn. 5, and Eqn. 6. The mode numbers $n$ are discrete integer indices and are paired to parameter values when finding solutions. The waveguide produces a spatially and spectrally varying pseudo-random "speckle" pattern described by the below equation.

Eqn. 7 $$P_Y(x,z=L,f) \propto \left|\sum_{n=1}^{m} E_Y(x,z=L,n,f)\right|^2$$

Here, the output speckle pattern results from the summation of all waveguide modes. The speckle pattern, $P_Y(x,f)$, defines the mixing matrix, $\Phi$, for compressive sensing. Fig. 2(a) shows the calculated speckle pattern at the output of a 100 μm wide, 11 cm long silicon waveguide when a narrow-band 1500 nm laser is input. The resultant speckle pattern demonstrates spatial pseudo-randomness sufficient to proceed with a waveguide design. While this calculation considers a straight waveguide, a spiral waveguide making use of waveguide bends is expected to generate a speckle pattern with similar or greater (favorable) spatial noncorrelation.

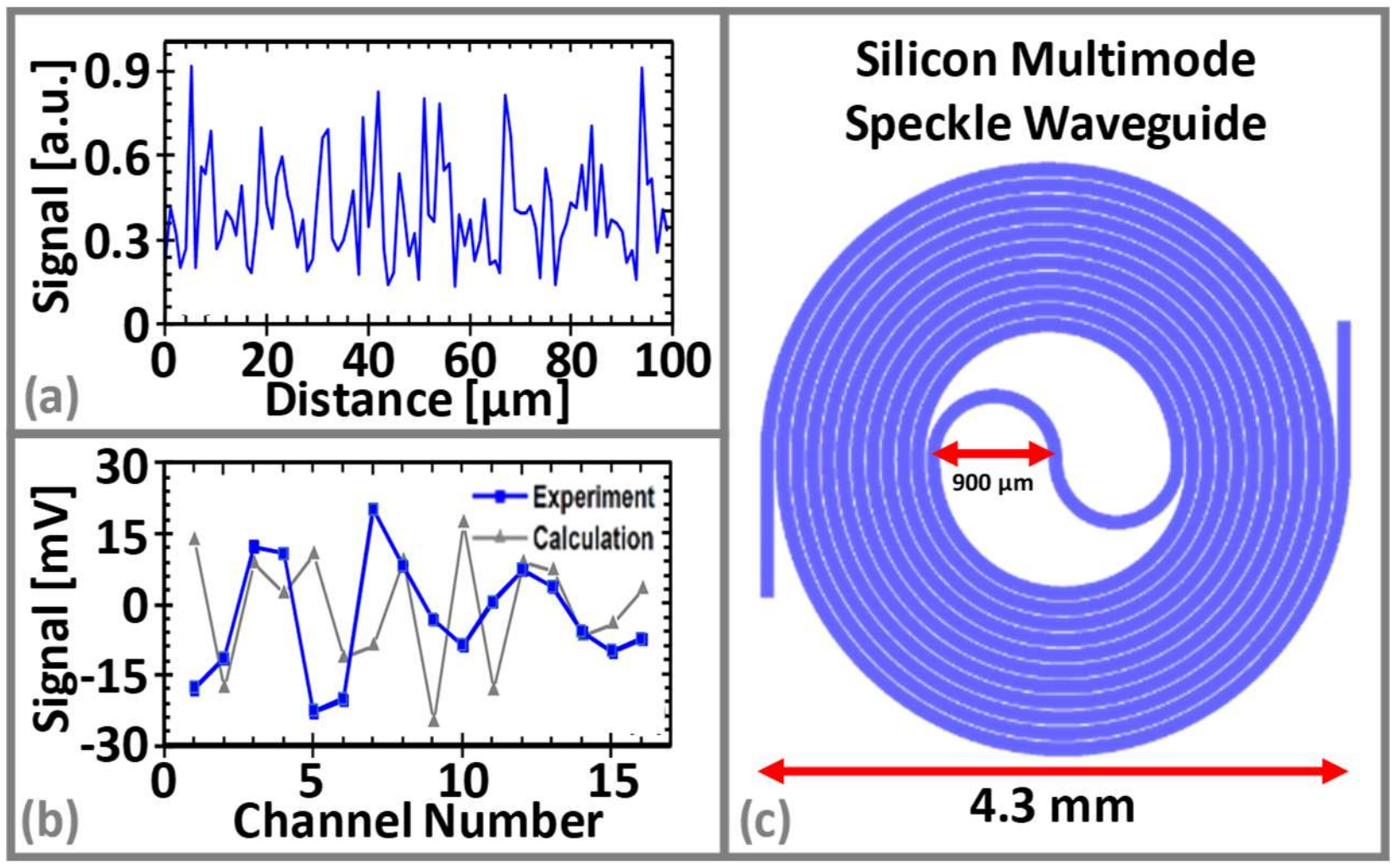

Fig. 2: (a) Calculated speckle pattern for a 11 cm long 100 µm wide silicon multimode speckle waveguide. (b) Experimentally measured speckle pattern from the 16 channel compressive sensing system described in this manuscript. Speckle pattern in (b) is resultant when no radio-frequency signal is applied to the compressive sensing system. (c) Rendering of photomask artwork showing the 10.915 cm long 96 µm wide multimode silicon speckle waveguide used in this work.

To conform to available die area, we designed a spiral waveguide. A rendering of the experimental waveguide's photomask artwork is shown in Fig. 2(c). The waveguide is 10.915 cm long, 96 µm wide, and 220 nm thick. The minimum spiral bend radius is 450 µm to reduce bend-loss in the multimode sections, and 15 µm separates each adjacent multimode waveguide. The spiral waveguide's inclusion of curves is favorably expected to generate additional speckle beyond that of the straight waveguide simulated in Fig. 2(a).

A multimode waveguide width of 96 µm is chosen primarily to enable sufficient speckle generation, per its proximity to the 100 µm simulations. Secondarily, a 96 µm multimode waveguide easily accommodates output spatial sampling of the speckle pattern. The multimode waveguide's output is spatially sampled via a 32-channel single mode waveguide bus, where each of the 32 waveguides is separated by a pitch of 3 µm.

An experimental validation of the waveguide's ability to produce speckle was performed. The input of the experimental waveguide was excited with narrow-band 1550 nm light. The 32 output channels were detected using balanced photodetectors to generate 16 channels for electrical digitization. The resulting experimental speckle pattern is shown in blue in Fig. 2(b). By comparison, the simulated data of Fig. 2(a) was numerically binned into 32 channels, differenced, rescaled, and plotted in gray in Fig. 2(b). Note that the pseudo-random nature of the theoretical and experimental speckle patterns are qualitatively similar, indicating the experimental waveguide produces sufficiently noncorrelated speckle. Quantitative agreement is not expected owing to the differences in the simulated design and the experimental design, as well as fabrication imperfections. Additionally, compressive sensing does not require one particular unique measurement matrix solution. The speckle-based measurement matrix must only exhibit a sufficient level of repeatable pseudo-randomness, making multimode speckle waveguides and their corresponding fabrication imperfections excellent candidates for analog-domain compressive sensing.

## 4. PIC Development

Harnessing waveguide speckle for use in compressive sensing of RF signals presents some unique challenges outside of routine photonic integrated circuit (PIC) techniques. Therefore, this section addresses PIC considerations that have been taken, which may be beyond the common practices of prevailing telecommunications and lidar PIC disciplines. However, the material presented may eventually find use in those disciplines. Considerations include: coupling the pulses from a mode-locked laser, which have a bandwidth of about 27 nm, to the PIC; development of a matched waveguide bus flare for the phase sensitive routing of a 32 channel waveguide bus; and a matched waveguide bus 90 degree bend for the phase sensitive routing of a 32 channel waveguide bus. Additional topics discussed in this section include: our primary PIC design and its fiber-array input/output configuration, a secondary design intended for surface coupling to a photodetector array, and our fiber-array PIC package. The PIC and fabrication process was designed at The Aerospace Corporation, fabricated by Freedom Photonics, and packaged by PLC Connections.

### *1. Input Grating Coupler*

The speckle mixer relies on an RF signal being imprinted on a broadband chirped pulse. Therefore, the ability of a broadband chirped pulse to couple to the PIC is critical to this work. This work uses chirped pulses centered about 1557 nm with a 27 nm bandwidth generated from a mode-locked laser (MLL) as previously utilized and reported in [3].

The well-known uniform pitch focusing grating coupler was modelled to accommodate the 27 nm bandwidth of this specific chirped pulse source. The resulting grating coupler utilizes a 628 nm pitch, 50% grating duty cycle, and a 38 degree included taper. The grating coupler is designed for an 8 degree launch angle when utilizing index matching fluid. The PIC is fabricated from a passive silicon-on-insulator wafer from SOITEC with 220 nm thick silicon, a 74 nm shallow etch, and a 220 nm through etch.

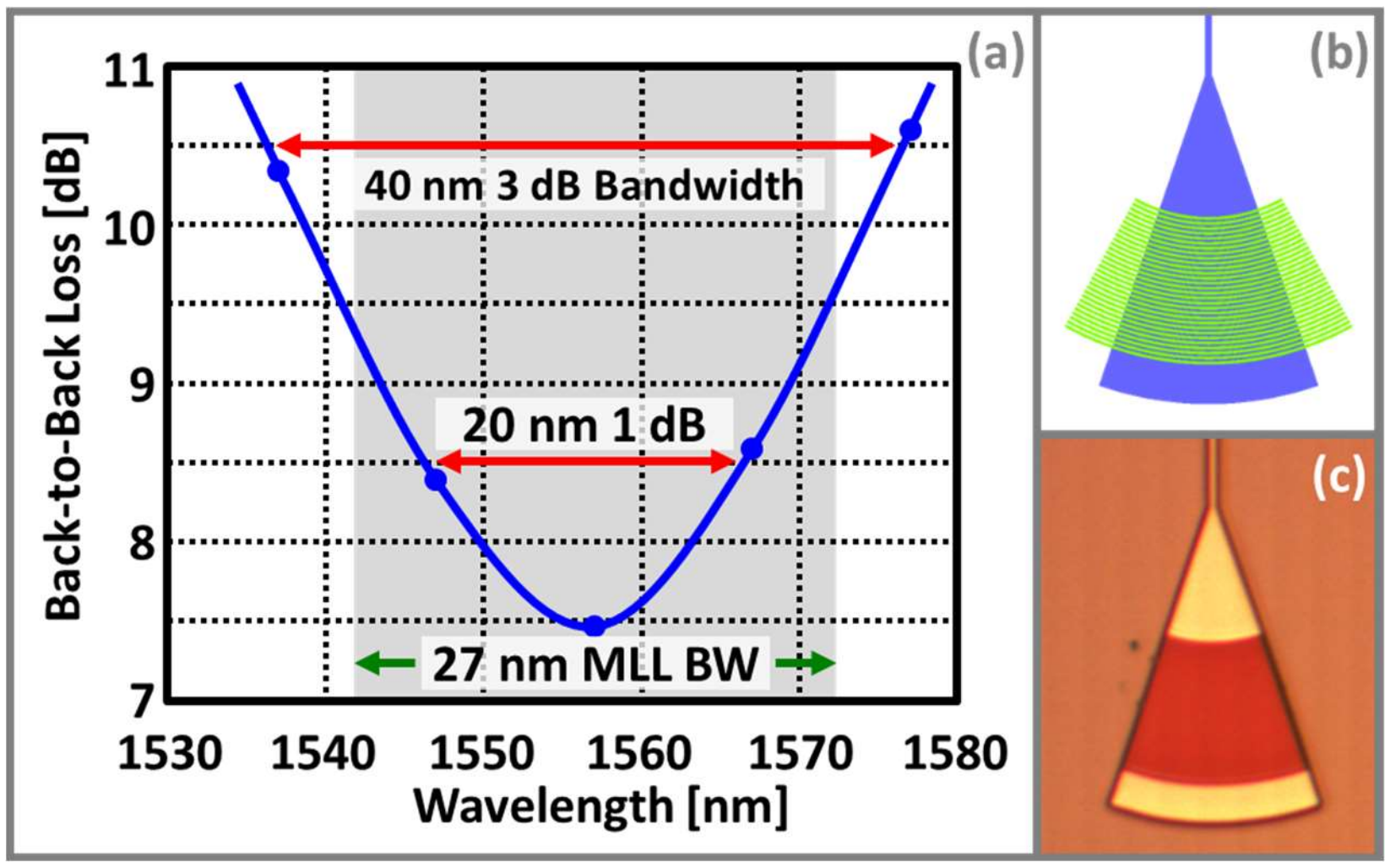

Fig. 3: (a) Measured back-to-back optical power loss for two grating couplers and a 1 mm length of single-mode waveguide as a function of wavelength, (b) grating coupler design artwork, and (c) grating coupler micrograph.

The grating coupler's artwork is shown in Fig. 3(b), and a micrograph of the fabricated device is shown in Fig. 3(c). The measured back-to-back optical power loss for two grating couplers and a 1 mm length of single-mode waveguide is shown in Fig. 3(a) as a function of wavelength. Ignoring the waveguide loss, each grating coupler exhibits an insertion loss of 3.75 dB. The pair of grating couplers demonstrate a 3 dB bandwidth of 40 nm, which is suitable for coupling the mode-locked laser into the PIC, and for extracting the speckle pattern output.

Note that grating couplers are highly polarization selective, and our MLL utilizes an unpolarized SMF-28 output fiber. Selection of a dual-polarization grating coupler or the use of a polarized MLL is recommended for future iteration to improve optical system link budget.

## 2. *Fiber Array PIC Design & Matched Waveguide Bus Flare*

The 6 x 6 mm design artwork shown in Fig. 4(b) supports a linear array of 44 optical couplers at 127 µm pitch. A 1x48 fiber array is used to interrogate the 44 optical couplers on the PIC. Two couplers are dedicated to over-head imaging beacons for first-light fiber array alignment. Four couplers serve two alignment loops for automated alignment. The PIC-based multimode waveguide was developed to serve as drag and drop replacement for the multimode fiber, imaging system, and 32-core fiber bundle in the experimental testbed demonstrated in [3]. The testbed's ADC bank supports 16 channels, where each channel is fed by two differentially detected optical signals. Therefore, the PIC was designed to support 32 output channels.

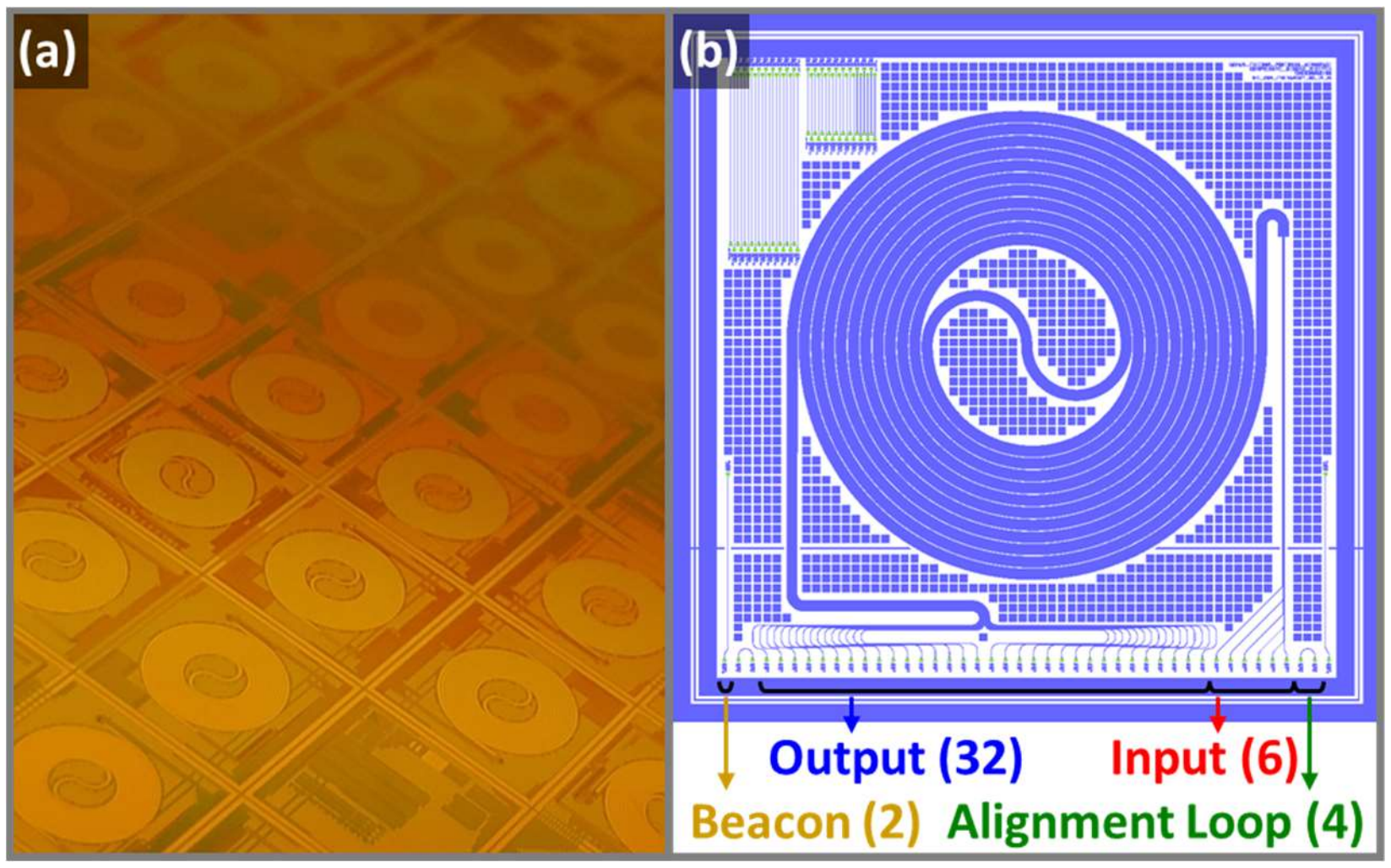

Fig. 4: (a) Micrograph of wafer-level photolithographic production of compressive sensing receivers, and (b) design artwork for the fiber-array compressive sensing receiver.

During testing, the PIC is fed by only a single input. However, 6 equivalent inputs were populated and feed the multimode waveguide. The purpose of the multiple input waveguides is to allow for optically induced input coupler damage during testing without rendering the entire PIC inoperable. 60 mW of optical power was coupled through the input couplers and no couplers were damaged. In addition to the approximately 11 cm long multimode speckle mixer introduced in Fig. 2(c), the PIC contains a number of test waveguides. Fig. 4(a) shows a micrograph of wafer-level photolithographic production of compressive sensing receivers.

The speckle mixer output is spatially sampled on the PIC into 32 single-mode waveguides, and each waveguide is routed to a fiber grating coupler. The spatial sampling results in a bus of 32 waveguides with a waveguide pitch of 3 µm. This waveguide bus is routed near the edge of the pic before being flared to produce a waveguide bus with a pitch of 127 µm. The 127 µm pitch mates with standard fiber arrays. To preserve the optical path length between each of the waveguides during the bus flare, a "waveguide bus trombone flare" is developed (c.f. Fig. 5).

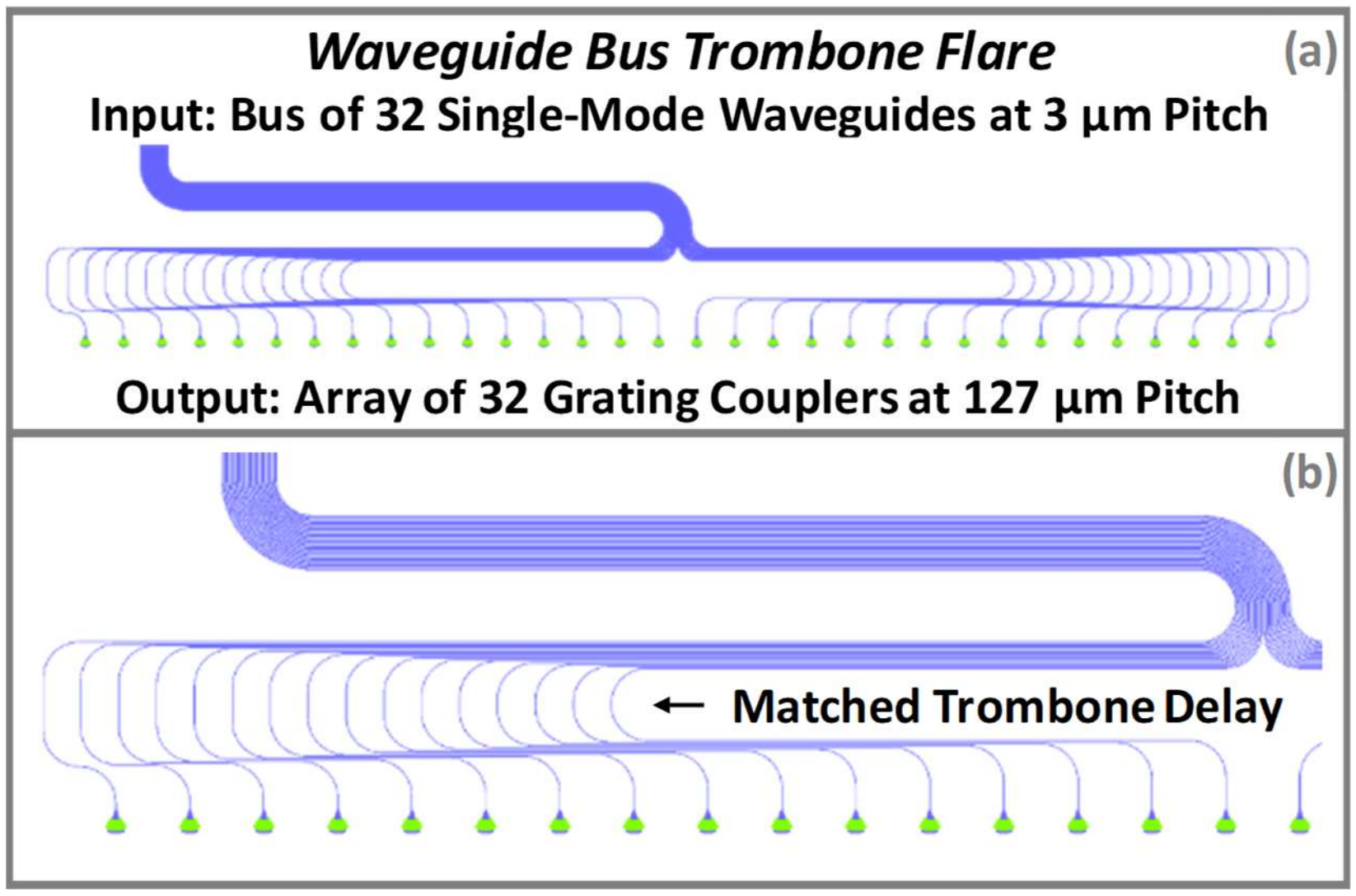

Fig. 5: A waveguide bus trombone flare is developed and utilized to maintain equivalent optical path delay between 32 waveguides of a waveguide bus when the inter-waveguide pitch is flared from 3 µm to 127 µm. (a) The full extent of the bus flare, (b) an enhanced view of the matched trombone delay, and (c) a highlight of the inner and outer waveguides with component straight waveguides labelled.

The matched-length trombone bus flare provides equivalent delay between all 32 waveguides (c.f. Fig. 5(a)). Each waveguide experiences four (4) of the exact same 90-degree bends and the exact same cumulative length of straight waveguide (c.f. Fig. 5(b)(c)). Fig. 5(c) has the same scale as Fig. 5(b), but only the inside and outside waveguides are drawn. In Fig. 5(c) the straight waveguides of the inner and outer trombone waveguides are labelled. Here $A + B + C + D + E = A' + B' + C' + D' + E'$. The waveguide on the extreme outside of the bus experiences the longest straight waveguide excursion (B) and the shortest straight waveguide return (D). The inner bus waveguides experience the shortest straight waveguide excursion (B') and the longest straight waveguide return (D'). Waveguide bends and straight waveguides typically exhibit dissimilar properties, resulting in: (1) unequal loss, (2) unmatched optical path lengths when physical lengths are matched, and (2) higher order dispersion effects. The matched-length trombone bus flare architecture provides a technology agnostic matched waveguide flare that is not influenced by the loss, optical path length, and dispersion mismatches between straight and bent waveguides. This is especially valuable for systems that require broadband optical signals such as mode-locked lasers and chirped pulses.

Because the footprint of this matched-length trombone bus flare is dependent on its constituent bend radii, it introduces 250 µm excess lateral footprint and approximately 295 µm of vertical footprint for the 62.5 µm radius bend technology shown here. Euler or other bends technology could equivalently be used in the developed waveguide bus trombone flare. The important feature is that each waveguide path experiences exactly the same number and type of bends, and that each waveguide path experiences the same length of straight waveguide. This ensures the final output signals are properly synchronized. The benefit of utilizing matched waveguide routing is dependent on the signal of interest and detection scheme. Path mismatch manifests in the RF photonic signal fidelity depending on sampling method and the allowable delay mismatch. The compressive sensing approach presented here, with a sample rate of 35 MHz, is exceptionally tolerant of path mismatch. Nevertheless, the waveguide bus trombone flare is expected to become an enabling component for phase sensitive and high-frequency analog optical signal processing that makes use of high channel count architectures. Applications include speckle enabled recurrent neural networks [5].

### 2. *Photodiode Array PIC Design & Matched 90 Degree Bus Bend*

A second PIC was designed to be co-packaged with two commercial photodiode arrays. Each photodiode array hosts 16 photodiodes on a 250 µm pitch. The PIC design facilitates photodiode packaging via the grating couplers on the west and east edges of the PIC (c.f. Fig. 6(a)). When completed, the co-packaged PIC and photodiode array assembly can be integrated with a custom ASIC for signal read out. The optical input mode is selected during attachment of the input fiber array to either an array of grating couplers or an array of end-fire couplers. As described in the previous section, loop back alignment aids ensure precision alignment, and only a single optical input is used during PIC operation. At the time of this article's submission the PIC has been designed, fabricated, and fibers have been attached. Development of the ASIC and the co-packaging subassembly is ongoing and may be reported in a future paper. System results reported in later sections utilize the PIC shown in Fig. 4.

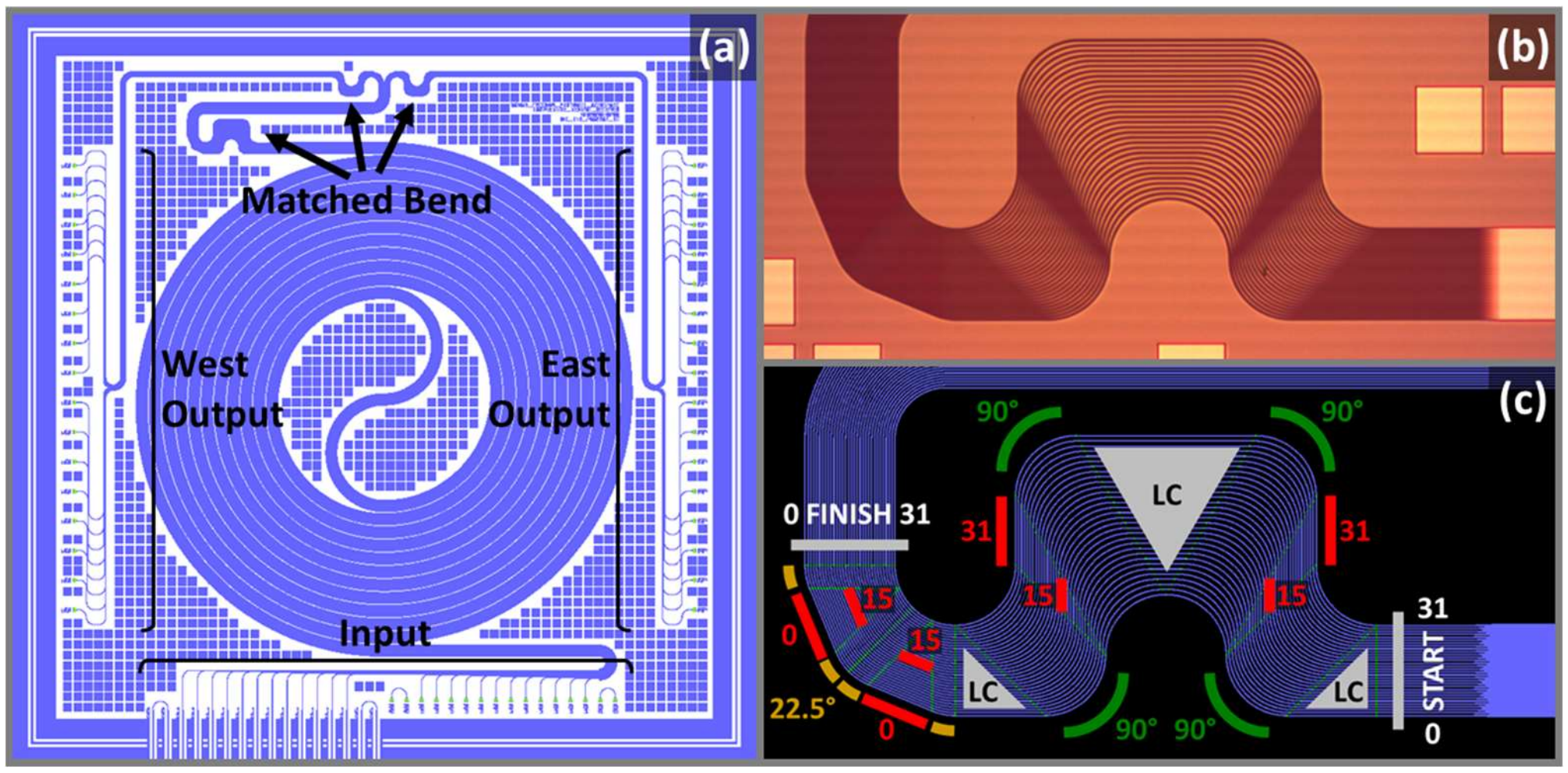

Fig. 6: A PIC is developed to excite two 1x16 commercial photodiode arrays using 250 µm pitch grating couplers (a). To maintain an equal path length between all 32 channels in the waveguide bus, a matched 90 degree bus bend is developed and fabricated (c) and (b), respectively.

The speckle pattern of the multimode waveguide is spatially sampled into 32 single mode waveguides at the top of the spiral waveguide, before the first matched bend, c.f. Fig. 6(a)(b). The transition between the 96 µm multimode waveguide and the 32 single mode waveguides is imaged on the lower right of Fig. 6(b). Note the lighter appearance of the multimode waveguide followed by the darker appearance of the single mode 32 waveguide bus. The PIC is designed to maintain an equal path length between all 32 output channels from the sampled speckle pattern to the output grating couplers. To accomplish this, a matched 90 degree bus bend is developed and fabricated as shown in Fig. 6(b)(c). Traditional 90 degree bus bends (unmatched) exhibit a different radius of curvature and optical path length for each individual waveguide (c.f. top left corner of Fig. 6(c)). The inclusion of two opposing traditional 90 degree bends (unmatched) cancels any optical path length difference to first order. Therefore, a matched 90 degree bus bend is needed primarily when there is an imbalance of opposing traditional bus bends. The principle of operation of the matched 90 degree bus bend is detailed in Fig. 6(c) and is inspired by the staggered starting positions of short distance track and field races. The matched bend begins at the "start" label and ends at the "finish" label. Each waveguide experiences four (4) of the exact same 90 degree bends (green annotation), four (4) of the exact same 22.5 degree bends (yellow annotation), and the exact same cumulative length of straight waveguide (red annotation). The lateral straight sections transport each waveguide the same distance laterally (from right to left) and are therefore equal. These sections are annotated "LC" in gray for lateral cancels, as the additional length experienced by these sections is automatically equal and is therefore matched. The individual waveguides are labelled 0 through 31. Waveguide 31 experiences the longest possible straight in the vertical direction going up and then coming down, and the shortest possible straight during the final bend (unlabeled). Note that the final bend accomplishes a 90 degree change in angle using four 22.5 degree bends and two lengths of straight waveguide. The 22.5 degree bends have the same radius for all waveguides. Conversely, waveguide 0 experiences the shortest possible straight in the vertical direction going up and then coming down (unlabeled), and the longest possible straight during the final bend. Waveguide 15 experiences a middle length during the vertical straights and the final bend. The straight waveguide lengths used in the final 90 degree bend are chosen to maintain the waveguide pitch between the start and finish of the final 90 degree bend. The lengths used in the vertical section are sequenced directly opposite those in the final bend, such that each waveguide experiences the same length of straight waveguide. In this way a matched 90 degree bus bend is accomplished that is waveguide technology agnostic where every waveguide experiences the same loss, optical path length, and dispersion. Similar to the waveguide bus trombone flare, the matched 90 degree bus bend is expected to become an enabling component for phase sensitive and high-frequency analog optical signal processing that makes use of high channel count architectures.

### *3. Fiber Attach and Packaging*

A 1x48 fiber array with a pitch of 127 µm was polished at an 8 degree angle and attached to the PIC by commercial vendor PLC Connections (c.f. Fig. 7). UV curable glue was used to adhere the fiber array to the PIC during active alignment of the loop back alignment structures. The PIC is suspended by the fiber array, and the fiber array is bedded in a 3D printed holder which protect the PIC from mechanical damage.

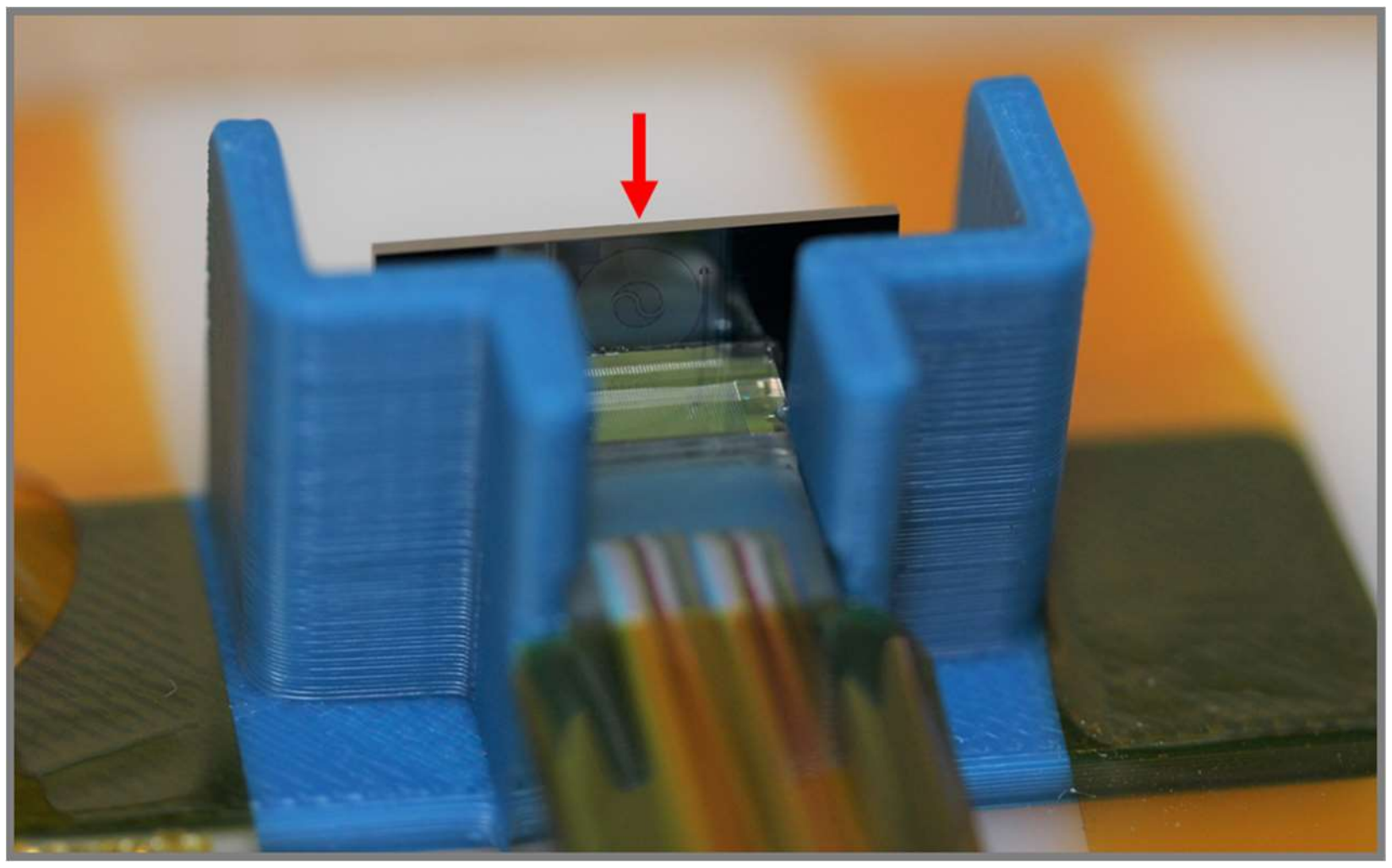

Fig. 7: The PIC is attached to a 1x48 V-groove fiber array using UV adhesive. The fiber array is bedded in a 3D printed holder, with the PIC suspended by the adhesive.

Note that the 11 cm long spiral waveguide is visible in Fig. 7 as indicated by the red arrow. The compressive sensing speckle mixer's insertion loss includes contributions from: 2 grating couplers, a 1:32 splitter, the 11 cm multimode waveguide, and approximately 1 cm of single mode routing waveguides. The insertion loss was measured by applying a tunable laser set to 1550 nm to the PIC input. A random output channel was selected and the insertion loss was calculated to be 36 dB. Future works may scan the laser wavelength to average insertion loss over wavelength. Averaging the insertion loss over all output ports was not performed, but should increase the insertion loss measurement accuracy. Proceeding with the 36 dB insertion loss measurement, the two grating couplers account for 7.5 dB, or 3.75 dB per grating coupler. The 1:32 split accounts for a 15 dB reduction in output channel power. The remaining 13.5 dB is accounted for by the 11 cm multimode waveguide and the 1 cm of single mode routing waveguides. Assuming the single mode waveguide and multimode waveguide exhibit the same loss, we measure a propagation loss of approximately 1.1 dB/cm which is consistent with literature results for reactive ion etched strip waveguides [19] [20]. Note that a shorter multimode waveguide with less loss may be sufficient to produce the required speckle. In addition, architectures making use of on-chip photodetection will both relax packaging considerations and drastically reduce the on-chip routing losses.

## 5. Experimental Setup

The experimental compressive sensing radio frequency receiver setup is shown in Fig. 8. The setup is the same as in [3] except that the multimode fiber, imaging apparatus, and 32-core fiber bundle in [3] have been replaced with the single PIC discussed in this work. Chirped optical pulses are created by passing femtosecond pulses from a mode-locked laser (35.71 MHz repetition rate, 27 nm optical bandwidth) through a spool of dispersion-compensating fiber (DCF) (-166 ps/nm at 1545 nm) producing 4.5-ns full-width half-max chirped pulses as shown within Fig. 8. In the DCF the blue-shifted light experiences a relatively greater phase velocity than the red-shifted light. This produces a well behaved linear relationship between time and wavelength in the pulses output from the DCF. RF signals from a 50 GSps arbitrary waveform generator (AWG) are intensity modulated onto the chirped optical pulses by a 1x2 lithium niobate Mach-Zehnder modulator (MZM) (zero-chirp, X-cut, 20 GHz 3-dB bandwidth) biased at quadrature.

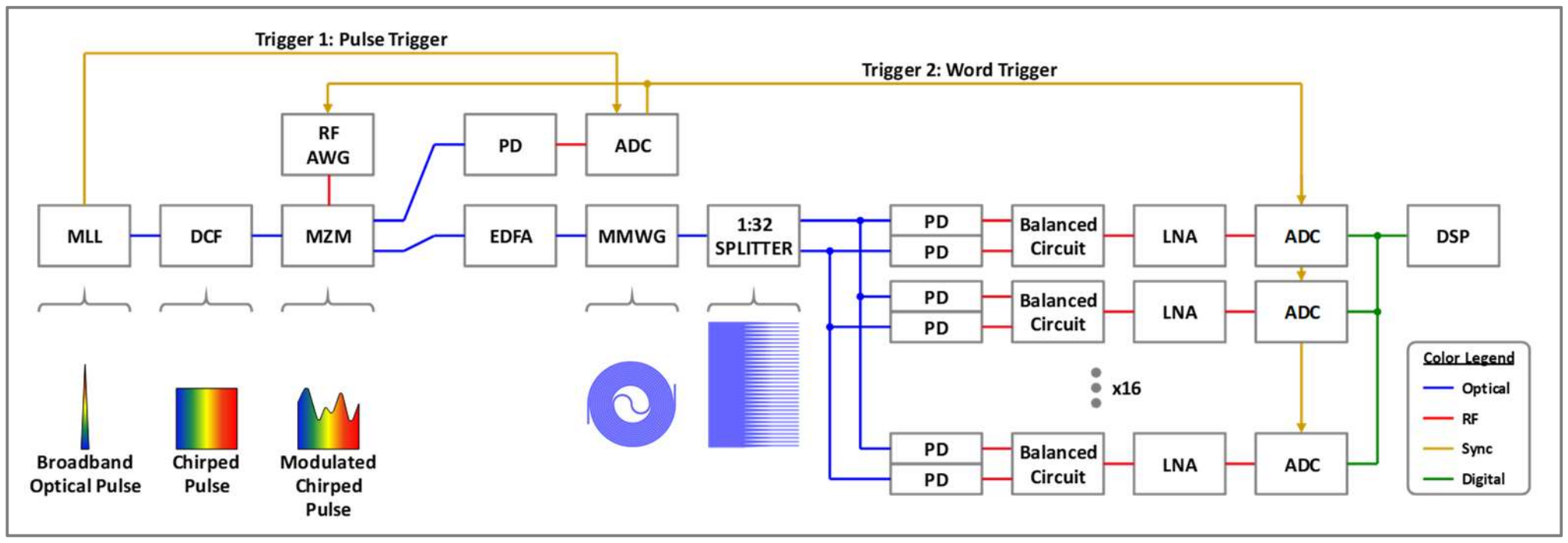

Fig. 8: Experimental setup. MLL: mode locked laser; DCF: dispersion compensating fiber; MZM: Mach Zehnder Modulator; EDFA: erbium doped fiber amplifier; MMWG: multimode waveguide; PD: photodiode; LNA: low noise amplifier; ADC: Analog to Digital Converter

One of the MZM outputs is monitored on an oscilloscope and used to set the MZM bias voltage, measure the MLL rep rate, and generate a word trigger for the RF AWG and ADC bank. The oscilloscope uses one pulse trigger from the MLL to generate a word trigger approximately every 500 µs or every 18k pulses to capture the full duration of the RF AWG word. The word trigger enables synchronous averaging of the RF AWG waveform.

The other MZM output is amplified by an EDFA to 60 mW average power before being input to the PIC. Note that PIC input damage is not expected below 300 or 450 mW, and 60 mW was the maximum power available experimentally from the EDFA used in this work. Increasing the input optical power may eliminate the need for the waveform averaging discussed later. The PIC hosts the multimode waveguide and the 1:32 splitter. In the multimode waveguide the dynamic RF signal and the static optical waveguide speckle mix to produce a dynamic output speckle pattern. The 1:32 splitter spatially samples the produced output speckle pattern into 32 bins. The matrix multiplication described in Eqn. 1 is thereby accomplish in the following manner: the RF signal mixing with the static speckle pattern, the speckle pattern being spatially sampled or binned, and sampled pulses being integrated over the pulse duration. This produces the final measurement signal vector $\vec{y}$, all components of which can be measured simultaneously. The analog equivalent of the entire compressive sensing matrix multiplication is thereby performed optically, in real time, with no digitization, manipulation, or storage of uncompressed signals.

The photodiodes, balanced circuits, LNAs, ADCs, and DSP discussed in [3] remain unchanged. The 32 single ended photodiodes are configured as differential pairs using a balancing circuit, the signal is amplified by a low noise amplifier, and digitized to yield a compressive sensing measurement matrix of 16 rows. Differential detection reduces the common mode noise and unmodulated pulse envelope contributions to the digitized signal. This enables more precise detection of the pertinent signal features using the limited 8 bit ADCs. After amplification by a LNA the 16 channels are digitized using a total of four (4) Tektronix DPO3034 4-channel oscilloscopes. Oscilloscopes were chosen for the experimental setup as the simplest solution for a multi-channel back-end digitizer satisfying sample-rate (35.71 MSps minimum based on the MLL rep-rate) and signal-conditioning (vertical offset, variable front-end amplification, etc.) requirements. Over-sampling by the oscilloscopes (2.5 GSps) enables the post-processing implementation of analog pulse integration and sample-and-hold modules while eliminating the need for accurate intra-pulse clock synchronization. Commercial systems routinely implement clock synchronization, analog pulse integration, and sample-and-hold solutions, in which case each channel would require a sample rate of 35.71 MSps.

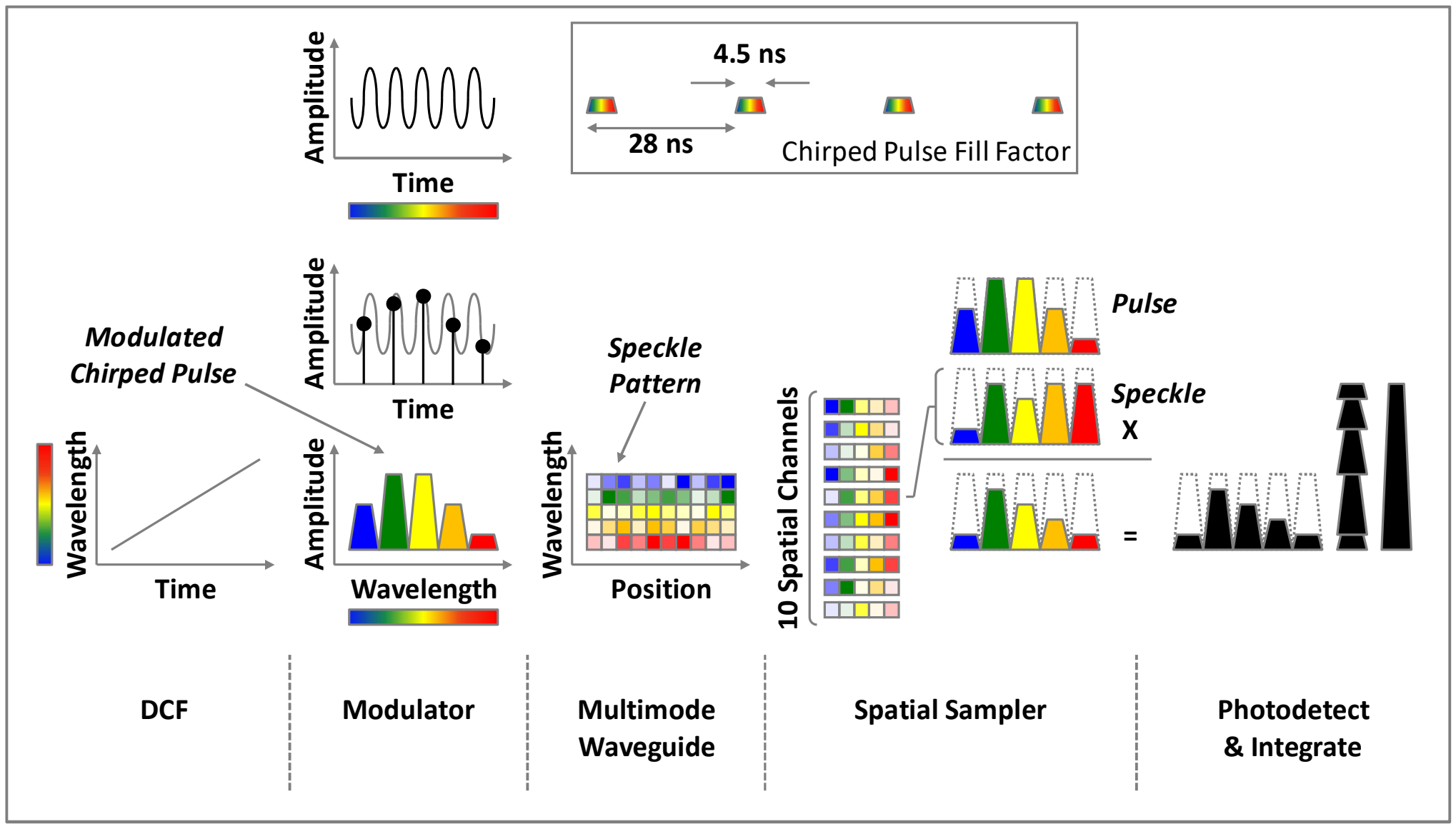

Fig. 9: Waveguide speckle and photodetection are utilized to perform row-column multiplication and summation, respectively, to accomplish matrix multiplication in the analog domain to enable compressive sensing (c.f. Fig. 1).

The experimental principle of operation is further detailed in Fig. 9. Starting on the left, the chirped pulse leaving the DCF fiber is plotted. Note that the time to wavelength mapping is linear. Second, the Modulator section is examined. On the top of Fig. 9 the input RF waveform is shown to vary with time. Below this, the RF waveform amplitude is sampled at 5 positions in time. Below this, the 5 positions in time are shown to correspond to 5 different wavelengths, and each of the wavelength amplitudes corresponds to the sampled RF signal amplitude. The result of is results a modulated chirped pulse. On the top right of Fig. 9, it is show that the mode lock laser produces a pulse once every 28 ns, but the pulse duration is only 4.5 ns. Therefore, 84 percent of the RF waveform is not sampled in time. However, it is not necessary to have full temporal coverage of the RF signal. Third, the chirped signal is input to the multimode waveguide. Here a conceptual wavelength dependent speckle pattern is plotted. The spatial or position axis is horizontal, and the wavelength axis is vertical. Note the dark colors denote minimal attenuation of that particular color, and so the color is bright. The light colors denote strong attenuation, and so the color is weak. Fourth, spatially sampling the speckle collapses the continuous wavelength dependence of propagating speckle into discrete channels. This is shown by transposing the previous conceptual speckle pattern and separating the patterns along the spatial axis. Examining one particular channel shows a wavelength dependent amplitude mask, labeled "Speckle". Here the blue is strongly attenuated, the yellow is moderately attenuated, and the green, orange, and red are weakly attenuated. The speckle mixer operates by multiplying the modulated chirped pulse, labeled "Pulse", by this wavelength dependence to yield the spatially sampled signal. Fifth, Photodetection collapses the wavelength dependence of the signal. Integration serves to sum each wavelength contribution. Invoking the compressive sensing introduction (c.f. Fig. 1), we can see that the "M" dimensions of the measurement vector "y" correspond with the number of spatial channels. The multiplication of the chirped pulse with the wavelength dependent speckle performs, in the analog optical domain, a single row-column multiplication between the measurement matrix "phi" (or the speckle) and the input signal vector "x" (or the RF waveform). The photodetection and integration performs the addition of each of the row-column results. This marks the final operation in matrix multiplication; row-column element wise multiplication followed by result addition. Note that integrate and dump circuitry is capable of performing the addition in the analog electrical domain without any imposing any frequency limitations on the incoming RF signal.

The calibration and signal reconstruction procedure discussed in [3] remain unchanged. To calibrate the system, the system input is excited by a sequence of RF tones that constitute a sparse signal dictionary over the entire frequency detection range. 50 to 100 laser pulse measurements are typically made per dictionary frequency. After the last dictionary tone, test signals to be recovered (multi-tone combinations of the dictionary frequencies) are input to the system. RF waveforms are generated using a high-speed arbitrary waveform generator (50 GS/s; Tektronix AWG70001A).

When the RF frequency is not an integer multiple of the laser repetition rate, the RF phase relative to the laser pulse drifts during the detection in a defined manner. Each subsequent pulse's phase is offset by $f_L/f_{RF}$, where $f_L$is the laser repetition rate and $f_{RF}$ is the particular RF dictionary frequency. In this way, the dictionary encodes both the phase and amplitude information for each dictionary frequency. In-phase and quadrature components of the measurement matrix can be calculated through singular value decomposition of the compressed signal for each dictionary frequency, or by fitting the compressed signal as a function of its phase offset to a sinusoid for each dictionary frequency, as described in [3]. Once the measurement matrix for the dictionary is calculated, the frequencies, phases and amplitudes of test signals can be reconstructed through $l_1$-norm minimization or orthogonal matching pursuit.

Signals with frequencies different from the dictionary frequencies, in general, can not be reconstructed through simple $l_1$-norm minimization but may be found through linear regression with an appropriate dictionary step size [21]. Therefore, for any real application, the dictionary must be chosen to span the frequencies that need to be recovered. For the demonstration in this work, up to 400 μs worth of RF information is overlaid on a series of pulses that are sampled in a single shot. The long RF signal generally contains multiple signals arranged in succession, each with fixed RF characteristics that lasts long enough to cover multiple laser pulses.

## 6. Results

The RF signal produced by the AWG consists of the dictionary followed by the test signal to be reconstructed, all of which is measured by the oscilloscopes in a single acquisition. Fig. 10 shows the acquired signals for four representative channels. The AWG is programed such that the dictionary and signal regions are separated by time intervals where there is no applied signal. This allows the RF background to be accurately calculated and subtracted, and the dictionary and signal regions to be easily parsed and separated using simple algorithms.

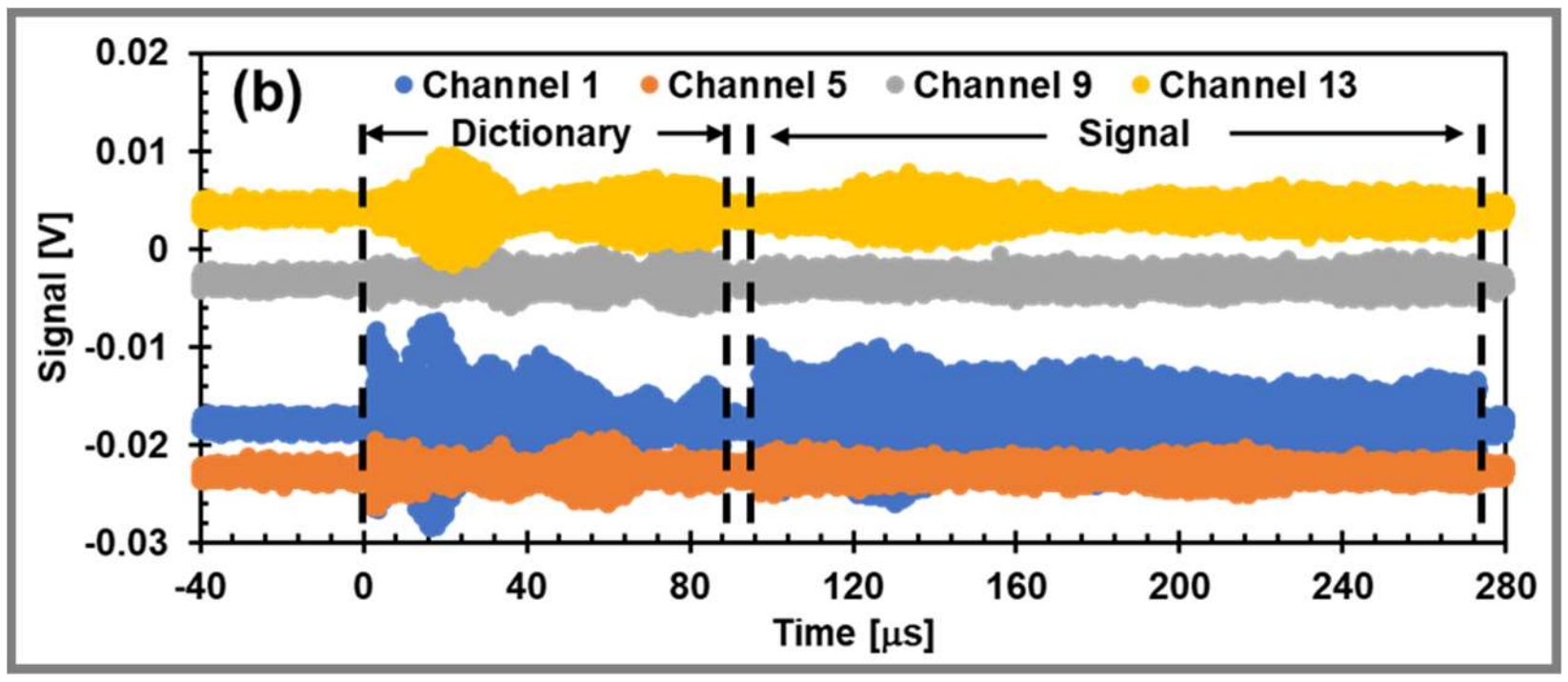

Fig. 10: Compressed signal as a function of time for the dictionary, from 0 to ~90 µs, followed by a two-tone signal from 96 µs to 270 µs.

This method allows system calibration to be performed regularly, within microseconds of signal capture, reducing the impacts of long-term drift due to heating, laser drift, and electrical fluctuations. Note that this calibration scheme was developed following experiments with a fiber based multimode waveguide. While such experiments have not yet been performed, it is theorized that integration of the setup into a rigid PIC will reduce some sources of drift to enable less frequency calibration. A benefit of the calibration scheme is the ability to quantify the reconstruction integrity by reconstructing the dictionary itself from the constructed measurement matrix.

Replacement of the multimode fiber in [3] with the multimode PIC-based waveguide resulted in additional system loss and reduced the system signal to noise ratio (SNR). Therefore, averaging over 512 identical signal sequences was necessary to reconstruct the RF signals. Averaging of 512 AWG waveforms proceeds as follows: a MLL pulse triggers the monitor ADC (c.f. Fig. 8), which generates a word trigger every 500 µs (approximately every 18k pulses) in order to trigger both the AWG and the oscilloscope bank. Using this trigger setup, jitter from the word trigger, AWG, and oscilloscopes limited measurements to RF frequencies below 4.5 GHz. Note that PIC input damage is not expected below 300 or 450 mW, and 60 mW was the maximum power available experimentally from the EDFA used in this work. Increasing the input optical power may eliminate the need for the waveform averaging using the current PIC design. Future PIC designs may be developed to improve SNR by directly integrating photodiodes on chip to avoid output grating coupler losses and output waveguide routing losses. Additionally, a decreased multimode waveguide length will reduce optical loss and may be suitable to achieve the required degree of speckle mixing.

To demonstrate compressive sensing and recovery, three different RF test signals were programmed into the AWG, sent through the system, digitized, and reconstructed. The test signals were reconstructed using the training dictionary only; no advanced knowledge of the test signals were passed to the reconstruction routines. A calibration dictionary consisting of 57 frequencies from 2.55 to 4.51 GHz, with a step size of 35 MHz, was used, with each dictionary frequency persisting for 60 laser pulses. Fig. 11(a) shows successful reconstruction of the in-phase and quadrature components of the training dictionary at the 57-frequencies. Patterns representing a sub-sampled sinusoid in the I/Q phase diagram represent successful reconstruction of the training dictionary, with the slope of the lines connecting subsequent points indicating the relative phase shift from peak to peak.

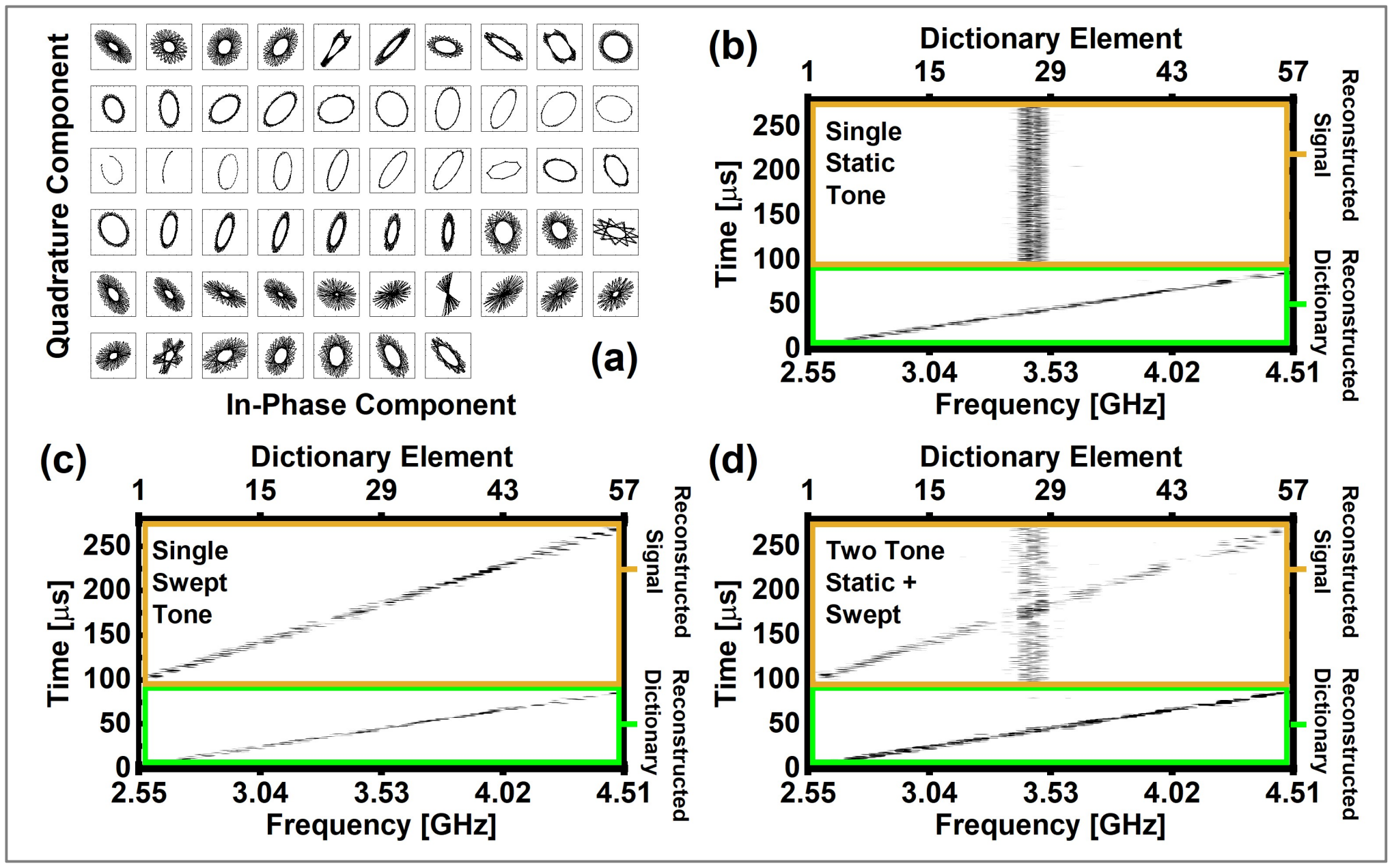

Fig. 11: (a) Successful recovery of dictionary signal phase (I & Q) using compressive sensing (2.55 GHz top-left, to 4.51 GHz bottom-right). (b) Reconstructed dictionary and static single-tone test. (c) Swept single-tone test. (d) dual-tone test with one static tone and one swept tone.

Fig. 11(a) shows that for each frequency in the dictionary the successfully recovered in-phase and quadrature points lie on a well-defined ellipse. The first test signal is a static single-tone test with frequency 3.5 GHz. Fig. 11(b) shows the reconstructed training dictionary used in the test (green box) followed by the successfully reconstructed static single-tone signal (yellow box). In this figure, the $l_1$-norm reconstructed signal power, consisting of the sum of the squares of the in-phase and quadrature components is plotted as the gray scale versus time and dictionary frequency. The second test signal is a swept single-tone test. In this test swept frequencies were selected to coincide with the dictionary frequencies from 2.55 to 4.51 GHz. Fig. 11(c) shows the successfully reconstructed swept single-tone signal. The third test signal is a two-tone test with one static tone and one swept tone. This test signal is essentially the sum of the first two test signals. Fig. 11(d) shows the successfully reconstructed dual-tone test signals. Fig. 11(d) demonstrates the important capability of this system to distinguish the multiple frequencies of a complicated signal. Note that, excepting waveform averaging to improve SNR, all of these demonstrations recover the signals of interest using real time signal capture during a 200 μs acquisition time.

Fig. 11(a) shows that dictionary phase can be successfully recovered with low noise. Fig. 11(b-d) plot the relative power of the $l_1$-norm recovered signal as functions of frequency and time for the three signal types discussed above. Power is concentrated in one frequency for the first signal type, along a sloped line for the second signal type, and in two distinct frequencies for the third type, just as the signals were input. This successfully demonstrates that speckle in a planar waveguide PIC can be used for compressive sensing of complex RF signals in the GHz band.

## 7. Comments on Future PIC Improvements

This work demonstrates the first use, to the authors' knowledge, of PIC-based waveguide speckle for the compressive sensing of RF signals, and therefore, there are many PIC improvements that can be made. Future improvements to the PIC are discussed and identified here.

Firstly, the impact of sidewall roughness on the speckle mixer's mixing strength is warranted. While sidewall roughness is unwanted from a loss perspective, it should increase modal mixing, resulting in a shorter waveguide being required to produce the appropriate speckle, which ultimately may reduce loss. An experimental study of the required speckle mixer designs is a basis for future work. Additionally, the use of a crystalline multimode speckle waveguide and its susceptibility or immunity to environmental drifting can be studied.

Secondly, on-chip photodetection will improve waveguide-photodiode mode-matching, eliminate off-chip coupling losses, reduce output waveguide routing losses, and improves system SNR. Further, on-chip photodetection reduces output channel count limitations imposed by the fiber array size and packaging considerations. By increasing the maximum allowable output channel count, through compact packaging, on-chip photodetection enables more finely resolved spatial sampling of the speckle pattern. Increased channel count may also improve reconstructed signal quality and allow for reconstruction of signals with reduced sparsity.

Third, high speed modulators are available in a number of PIC material platforms and may further reduce overall system size and passive insertion losses. The performance of silicon and indium phosphide modulators with regard to analog frequency response, SNR, and spur-free dynamic-range (SFDR) may be inferior to more common microwave photonics materials such lithium niobate. However, silicon and InP modulators have demonstrated impressive performance metrics. The ability to integrate the modulator with other components to create a monolithic system likely outweighs any potential component performance reductions.

Finally, a chirped waveguide Bragg grating (WBG) could potentially be developed in a low-loss material system such a SiN to replace the DCF fiber. The feasibility of such a replacement may rely on the use of higher repetition rate lasers depending on the performance of the WBG. Active materials such as heterogeneous indium phosphide on silicon may enable the integration of the MLL on-chip. Active components enable the replacement of the EDFA with an on chip semiconductor optical amplifier (SOA) before the modulator (c.f. Fig. 8). An SOA on each output single mode waveguide channel following the speckle pattern spatial sampler could improve SNR. Increasing the resolution of the spatial sampling may enable increased frequency resolution. However, increasing the spatial sampler channel count increases the splitting loss, reducing SNR. SOAs on the output waveguide would counteract this scaling trend to improve the link budget and frequency resolution simultaneously.

## 8. Conclusion

The use of waveguide speckle for the compressive sensing of RF signals in a compact PIC was demonstrated for the first time. The demonstration serves to identify a pathway toward a fully integrated optoelectronic speckle-based RF compressive sensing system.

A formalism for quantifying the speckle produced by a PIC-based multimode waveguide was introduced, demonstrated, and utilized to design a multimode speckle waveguide. The constituent components needed to realize a speckle-based RF compressive sensing PIC were developed and described, including a MLL matched grating coupler, a matched trombone bus flare, and a matched 90 degree bus bend. A complete 6x6 mm compressive sensing PIC was designed, fabricated, packaged, and tested. The experimental test setup was refined to enable waveform averaging. The setup was used to successfully capture and reconstruct 2 GHz of RF signals from 2.5 to 4.5 GHz using a 35 MHz rep rate laser. The demonstrated approach reduced the required sampling rate by a factor of 114 compared with Nyquist sampling. Potential improvements to the PIC for follow-on work were discussed, including the use of: on chip photodiodes, integrated modulators, a waveguide Bragg grating, SOAs, and an integrated MLL. These improvements may enable improved frequency coverage and improve the system signal-to-noise ratio in compact form factor.

## Acknowledgment

This work was supported by NAVAIR Contract Number N68335-18-C-0089. A portion of this work was performed in the UCSB Nanofabrication Facility, an open access laboratory.